\documentstyle[12pt]{article}
\newcommand{\II}{{\cal I}}
\newcommand{\MM}{{\cal M}}

\newcommand{\wt}{\widetilde}
\newcommand{\tu}{\widetilde u}
\newcommand{\tv}{\widetilde v}
\newcommand{\tf}{\widetilde f}
\newcommand{\tg}{\widetilde g}

\newcommand{\be}{\begin{equation}}
\newcommand{\ee}{\end{equation}}
\newcommand{\ben}{\begin{eqnarray}\displaystyle}
\newcommand{\een}{\end{eqnarray}}
\newcommand{\refb}[1]{(\ref{#1})}

\newcommand{\sectiono}[1]{\section{#1}\setcounter{equation}{0}}

\begin{document}

{}~ \hfill\vbox{\hbox{hep-th/9611186}\hbox{MRI-PHY/96-33}}\break

\vskip 3.5cm

\centerline{\large \bf A Non-perturbative Description of the}
\centerline{\large \bf Gimon-Polchinski Orientifold}

\vspace*{6.0ex}

\centerline{\large \rm Ashoke Sen\footnote{On leave of absence from 
Tata Institute of Fundamental Research, Homi Bhabha Road, 
Bombay 400005, INDIA}
\footnote{E-mail: sen@mri.ernet.in, sen@theory.tifr.res.in}}

\vspace*{1.5ex}

\centerline{\large \it Mehta Research Institute of Mathematics}
 \centerline{\large \it and Mathematical Physics}

\centerline{\large \it 10 Kasturba Gandhi Marg, 
Allahabad 211002, INDIA}

\vspace*{4.5ex}

\centerline {\bf Abstract}

A T-dual version of the Gimon-Polchinski orientifold can be described
by a configuration of intersecting Dirichlet seven branes and orientifold
seven planes in the classical limit. 
We study modification of this background due to quantum corrections.
It is shown that non-perturbative effects
split each orientifold plane into a pair of
nearly parallel seven branes. Furthermore, a pair of intersecting 
orientifold planes, instead of giving rise to two pairs of intersecting
seven branes, gives just one pair of seven branes, each representing
a pair of nearly orthogonal seven branes smoothly joined to each other
near the would be intersection point. Interpretation of these results from 
the point of view of the dynamics on a three brane probe is also discussed.

\vfill \eject

\baselineskip=18pt

\sectiono{Introduction and Summary} \label{s1}

Orientifolds\cite{ORIENT,DBRANE} are generalization of orbifolds where the
group by which we mod out the theory includes a world-sheet parity
transformation, possibly in conjunction with some other internal and / or
space-time symmetry transformation. 
These models have been of interest lately since
they give rise to many new classes of string compactifications, some of
which are related to more conventional string compactifications by 
strong-weak coupling duality transformation. In particular, type I
theory, which can be viewed as the quotient of type IIB theory by the
world-sheet parity transformation, has been conjectured to be dual to
SO(32) heterotic string theory\cite{WITTEND}.

Much can be learned about orientifolds by compactifying type I theory on
a torus and then making $R\to (1/R)$ duality 
transformation on some / all directions of
the torus. Under this T-duality transformation, the world-sheet parity
transformation gets converted to a new transformation which is a 
combination of the world-sheet parity transformation, change of sign of
some of the space coordinates, and possibly some internal symmetry 
transformation. The hyperplanes in space time which are left invariant
by the space-time part of the transformation are known as orientifold
planes. These orientifold planes turn out to carry charge under a gauge
field originating in the Ramond-Ramond (RR) sector of 
the theory. In order to cancel this charge, we need to place appropriate
number of Dirichlet branes\cite{DBRANE} (D-branes) 
$-$ which are also known to carry
RR charges\cite{POLCHI} $-$ parallel to the orientifold planes.

A special class of orientifolds, obtained by compactifying type I theory
on a two dimensional torus, and then performing $R\to(1/R)$ duality
transformation in both circles, was analysed in detail in ref.\cite{SENF}.
This T-dual version can be obtained by modding out type IIB theory on
$T^2$ by a combination of the world-sheet parity transformation, an
internal symmetry transformation that changes the sign of all 
Ramond sector states
in the left moving sector of the string world-sheet, and a space-time
transformation that changes the sign of both coordinates of the torus.
The resulting theory contains four orientifold seven planes transverse
to the torus directions, each carrying $-4$ units of RR charge, and sixteen
Dirichlet seven branes parallel to them cancelling their RR charges.
It was shown in ref.\cite{SENF} that while this picture is valid in the 
classical limit, and also to all orders in the open string perturbation 
theory, non-perturbative corrections change this picture. In
particular, each orientifold plane splits into two seven branes, which
are related to the ordinary Dirichlet seven branes by SL(2,Z)
transformation of the type IIB theory. 

In \cite{GIMPOL} Gimon and Polchinski constructed a more complicated
class of orientifold
models describing N=1 supersymmetric theories in six dimensions. 
(See also ref.\cite{BIAN} for an earlier construction of these models
in a different formalism.)
These theories are obtained by compactifying type IIB theory on a 
four dimensional torus, 
and modding out the theory by a $Z_2\times Z_2$ symmetry, with the first
$Z_2$  generated by the world-sheet parity transformation, and the
second $Z_2$ generated by a geometric transformation that changes the
sign of all the coordinates of $T^4$ (which we shall denote 
by $x^6, \ldots x^9$). The result is
a configuration of orientifold five planes and nine planes (filling up 
the whole space-time), and Dirichlet five branes and nine branes.
We shall analyse a T-dual version of this model, obtained by making an
$R\to(1/R)$ duality in two of the coordinates of the torus (which we shall
take to be $x^6$ and $x^7$ for definiteness). The effect of
this T-duality transformation is to make the orientifold nine plane and
the Dirichlet nine branes into orientifold seven planes and Dirichlet 
seven branes transverse to $x^6$ and $x^7$. On the other hand the
orientifold five planes and Dirichlet five branes are also transformed
into orientifold seven planes and Dirichlet seven branes respectively,
but these are transverse to $x^8$ and $x^9$.
Thus the result is a set of intersecting
orientifold seven planes and Dirichlet seven branes.

The question that we ask is again, how does the non-perturbative effects
in the orientifold theory modify this picture? For a class of 
Gimon-Polchinski  models we are able to answer this question by 
analysing various consistency requirements. The basic idea, as in 
ref.\cite{SENF},is to use these consistency conditions to
determine the type IIB string coupling (the axion-dilaton modulus) as a
function of the space-time coordinates. This in turn gives us the locations
of the seven branes in space-time.  We find that these
corrections do not modify the configuration of intersecting Dirichlet
seven branes, but it does modify the configuration of orientifold 
planes. First of all, as in ref.\cite{SENF} each of the orientifold planes
splits into two seven branes. Thus naively we would expect that a pair of
intersecting orientifold planes will be described by two pairs of
intersecting seven branes. What we find instead is that a pair
of such seven branes, which are asymptotically orthogonal to each other,
now join smoothly near the expected point of 
intersection\footnote{Throughout the paper, when we refer to the `point of
intersection' of two seven branes, we really mean a five dimensional
submanifold.} to give one
smooth seven brane. Thus a pair of intersecting orientifold planes gets
transformed to pair of seven branes. In the general case these two
seven branes further join smoothly at their would be intersection points
to give a single seven brane; but in special cases they can remain as
two distinct seven branes.

Our results can also be interpreted as giving information about the
non-perturbative dynamics on a three brane probe in this orientifold
background\cite{BDS}. 
In particular it gives the U(1) gauge coupling 
on this three brane probe as a function of various moduli.
At the classical level the N=1 supersymmetric
world volume field theory on
this three brane probe has some novel features. For example, at a generic
point in the moduli space, there is a single U(1) gauge group, which
gets enhanced to SU(2) in different regions of the moduli space. If
we bring this three brane near the point of intersection of the two 
orientifold planes, then in the classical limit the three brane world
volume theory seems to contain two
different SU(2) gauge groups sharing this single U(1) group. 
The classical Lagrangian for this system, based on an $SU(2)\times SU(2)$
gauge group, can be constructed along the lines of ref.\cite{COSH}.
Once quantum
corrections are taken into account, however, 
the $SU(2)\times SU(2)$ symmetry is not restored near the 
region of intersection of the two orientifold planes.
Instead, in the infrared limit, the low energy theory
is an N=2 superconformal field theory with U(1) vector multiplet, and a 
massless charged hypermultiplet.

The rest of the paper is organised as follows. In section \ref{s2} we
describe the T-dual version of the Gimon-Polchinski model that will be
the focus of our attention. This gives a description of the background
in the classical limit. In section \ref{s3} we show how non-perturbative
quantum corrections modify this picture.

\sectiono{A T-Dual Version of Gimon-Polchinski Orientifold} \label{s2}

The Gimon-Polchinski orientifold\cite{GIMPOL} is obtained by
modding out type IIB string theory on $T^4$ by a $Z_2\times Z_2$
symmetry. If we denote by $\Omega$ the world-sheet parity transformation,
and by $\II_{6789}$ the transformation that changes the sign of all
four coordinates $x^6,x^7,x^8,x^9$ labelling the torus, then the first
$Z_2$ is generated by $\Omega$ and the second $Z_2$ is generated by
$\II_{6789}\cdot\Omega$. If we now perform an $R\to(1/R)$ duality
transformation in $x^6$ and $x^7$ directions 
then the generators of the two $Z_2$ transformations
get transformed to 
\be \label{en1}
g \equiv \II_{67}\cdot(-1)^{F_L}\cdot\Omega, \qquad \hbox{and} \qquad
h\equiv \II_{89}\cdot(-1)^{F_L}\cdot\Omega,
\ee
respectively, 
where $\II_{mn}$ denotes the transformation $x^m\to - x^m$,
$x^n\to -x^n$, and
$(-1)^{F_L}$ is the transformation that changes the sign of all the
Ramond sector states of the left-moving sector of the
string world sheet without affecting the right-moving and/or Neveu-Schwarz
sector states. 
This is the model on which we shall focus our attention. Note
that this description only specifies the action of the $Z_2$
transformation on the untwisted sector closed string states. Different
models have been constructed by exploiting the ambiguity in the action
of the $Z_2$ transformations on the twisted sector and open string 
states\cite{POLNEW,BLUZAF,DABPAR} (see also \cite{GIMJOH,GOPMUK}),
but for us the action of the $Z_2$ transformation on all the states
is fixed completely by demanding that this is the T-dual of the model
discussed in ref.\cite{GIMPOL}. We shall concentrate on the sector
of the theory that is connected to the $U(16)\times U(16)$ point, {\it i.e.}
the sector with no half five-branes in the language of 
refs.\cite{GIMPOL,WITSIX}.

To start with, let us set up some notations. We define complex coordinates
on $T^4$
\be \label{e0}
w = x^6 + i x^7, \qquad \qquad z = x^8 + i x^9.
\ee
These coordinates change sign under $g$ and $h$
respectively. We also introduce coordinates
\be \label{e11a}
u = w^2, \qquad \qquad v=z^2,
\ee
which are single valued on the orientifold. 

The theory described above has $N=1$ supersymmetry in six dimensions.
The spectrum of massless twisted 
sector / open string states in this
theory, as determined in ref.\cite{GIMPOL}, is as follows:

\begin{enumerate}

\item For each of the sixteen points on the torus fixed under $gh$ we get a
hypermultiplet of the $N=1$ supersymmetry algebra from closed string
states twisted under $gh$.

\item Due to the modding out by $g$,
each of the four orientifold seven planes fixed under $\II_{67}$
carries $-4$ units of charge under an RR gauge field.
This charge is cancelled by putting 16 ($=4\times 4$) Dirichlet seven
branes transverse to the 67 plane. This guarantees that the RR charge
is neutralised globally. It is neutralised locally when the 16
D- seven branes are grouped into four groups of four each, and each group
is localised at an orientifold plane. In order to describe the spectrum of
massless open string states for such a configuration, 
it is best to focus on one of the orientifold planes, since
each of them gives identical spectrum. If we ignore the projection by
the group element $h$, then the open string states starting and ending on
these D-branes and their images under $g$ give rise to $SO(8)$ gauge
fields, and a complex scalar (representing the motion of the seven brane in
two transverse directions) in the adjoint representation of
$SO(8)$. However, projection by $h$ acts on this gauge group as a gauge 
transformation
\be \label{e1}
\MM=\pmatrix{M_4 & 0 \cr 0 & - M_4}\, ,
\ee
where
\be \label{e2}
M_{2n} = \pmatrix{0 & I_n\cr -I_n & 0 \cr}\, ,
\ee
$I_n$ being the $n\times n$ identity matrix. Note that our notation differs
from that of \cite{GIMPOL,WITSIX} by a simple rearrangement of basis.
This breaks $SO(8)$ to a group that commutes with \refb{e1}, {\it i.e.}
to $U(4)$. We shall denote this group by $U(4)_v$ since it lives on the
seven branes transverse to the $u$-plane, {\it i.e.} parallel to the
$v$-plane. One loop anomaly effect breaks this further to 
$SU(4)_v$\cite{WITSIX}, but 
we shall continue to refer to the corresponding gauge
groups by $U(4)_v$ rather than $SU(4)_v$ keeping in mind that we are 
refering to the unbroken gauge symmetry {\it in the classical limit.}

The action of $h$ on the adjoint complex
scalar fields (which we shall denote
by $\phi_v$) living on these seven branes takes the form
\be \label{e7}
\phi_v (z) \to - \MM \phi_v(-z) \MM^{-1}\, .
\ee
For $z(\equiv\sqrt{v})$ independent $\phi_v$, 
this gives two complex massless scalars in
the ${\bf 6}$ representation of $SU(4)_v$, carrying $U(1)_v$ charge $\pm 1$.
These complex scalars,
together with the components of the $U(4)_v$ gauge fields along the 
$v$ plane which survive the $h$ projection, 
and the fermionic open string states, give rise to
two hypermultiplets of the N=1 supersymmetry algebra in six dimensions.

We shall consider breaking the gauge group further by giving vacuum 
expectation value (vev) to $\phi_v$ of the form:
\be \label{e5}
\langle \phi_v \rangle = \pmatrix{0 & {\bf m_d}\cr -{\bf m_d} & 0\cr}
\ee
where
\be \label{e6}
{\bf m_d} = \pmatrix{m_1 &&&\cr & m_2 &&\cr && m_3 &\cr &&& m_4}.
\ee
Physically this corresponds to moving the four seven branes away from the
orientifold plane, with $\pm m_i$ denoting the locations of the seven
branes and their images under $g$
in the $w$ plane. The projection \refb{e7} now requires
\be \label{e10}
m_3 = m_1, \qquad \qquad m_4=m_2.
\ee
In other words, the four seven branes, instead of being allowed to
move freely, can move only in pairs.\footnote{Actually, if 
we take into account
possible $v$ dependence of $\phi_v$, we see that the projection
\refb{e7} only requires a pair of seven branes to be joined togeter at
$v=0$. But if we want the seven branes to be parallel, this requires
that they must coincide for all $v$.} This vev breaks the $U(4)_v$
gauge group to $SU(2)_v\times SU(2)_v'$. A special case of this is
$m_1=m_2$; in this case the unbroken gauge group is $Sp(4)_v$.

As has already been pointed out, due to the presence of four
orientifold planes transverse to $u$ plane, the above story is 
repeated four times. Thus, for example, in the case where 
four seven branes (and their images) sit at each orientifold plane,
the net gauge group arising from the D-branes transverse
to the $u$-plane is $(U(4)_v)^4$.

\item The same story is repeated for the orientifold plane and the seven 
branes transverse to the $v$ plane. Thus, for example, each of the four
orientifold planes fixed under $\II_{89}$ carries $-4$ units of RR charge
which is cancelled by placing 16 D- seven branes parallel to these
orientifold planes. The charge is neutralised locally when 
each orientifold plane has four D-branes on top of it. In this case the gauge
group associated with each orientifold plane is again $U(4)$, which we
shall denote by $U(4)_u$. This can be broken to $SU(2)_u\times SU(2)'_u$ 
by pulling the seven branes away from the orientifold planes in pairs, which
can be interpreted as due to the vev of two hypermultiplets in the  {\bf 6}
representation of $SU(4)_u$. Again a special case of this, when the four
D-branes coincide, but are away from the orientifold plane, is $Sp(4)_u$.

\item Finally there are open string states starting on a seven brane 
parallel to the $u$ plane and ending on a seven brane parallel to the
$v$ plane. At the $(U(4)_v)^4\times (U(4)_u)^4$ point these hypermultiplets
transform in the $(4,4)$ representation for each pair of
$(U(4)_v, U(4)_u)$. By giving vev to these hypermultiplets it is
possible to break the gauge group completely, but we shall not discuss
this branch of the vacuum configuration in much detail.

\end{enumerate}

In order to simplify our analysis, we shall focus on the physics of only one
orientifold plane (together with four D-branes) parallel to the
$u$ plane intersecting only one orientifold plane (together with four
D-branes) parallel to the $v$ plane. This is in the spirit of the
analysis of ref.\cite{SENF} and captures much of the essential physics of
the problem. Such a configuration is obtained if we do not compactify the
type IIB theory on $T^4$ but simply mod out the type IIB theory in
(9+1) dimensional flat space-time by the $Z_2\times Z_2$ symmetry generated
by $g$ and $h$, and put appropriate D-branes to neutralise all RR charges
globally. In this theory the maximal gauge symmetry group in the classical
limit is $U(4)_v\times U(4)_u$. This can be broken to 
$SU(2)_v\times SU(2)_v'\times SU(2)_u\times SU(2)_u'$ by expectation values 
of complex scalar fields in the ${\bf 6}$ representation of both $SU(4)$'s.
Four complex parameters $m_1, m_2, m_1', m_2'$ characterise the vacuum 
expectation values of these fields. In the special case
$m_1=m_2$, $SU(2)_v\times SU(2)_v'$ gets enhanced to $Sp(4)_v$. Similarly
for $m_1'=m_2'$, $SU(2)_u\times SU(2)_u'$ gets enhanced to $Sp(4)_u$.

Besides these four complex parameters, we need a few others to
completely characterise the vacuum.
One of them is the asymptotic value $\tau$ of the
axion-dilaton field:
\be \label{e11b}
\lambda = \phi + i e^{-\Phi/2}\, ,
\ee
where $\phi$ denotes the RR scalar field of the type IIB theory and $\Phi$
denotes the dilaton field. Thus
\be \label{en2}
\tau = \lim_{u\to \infty \atop v\to \infty} \lambda(u,v)\, .
\ee
We set the asymptotic metric to be $\eta_{\mu\nu}$ by using the freedom
of general coordinate transformation in the (9+1) dimensional theory; thus
we do not get any extra parameter from the metric.
The only other parameters (besides vev of hypermultiplets in the (4,4)
representation of $U(4)_v\times U(4)_u$ which we are setting to zero) 
are the ones associated with the vev of
massless closed string fields originating 
in the sector twisted by $gh$. These 
correspond to the blow up modes of the orbifold singularilty at the
intersection of the two orientifold planes, which we shall take to be
the point ($u=0$, $v=0$). 
However, as was shown in ref.\cite{WITSIX}, due to anomaly effects
these modes acquire mass at one loop order if the unbroken gauge group
at the classical level contains at least one $U(1)$ factor. Thus we expect
these modes to be present only if both the $U(1)$ factors are broken at the
classical level by Higgs mechanism. In particular if $m_1=m_2=0$ or if
$m_1'=m_2'=0$, then these modes should be absent.

\sectiono{Non-perturbative Description of the Model} \label{s3}

\subsection{The Problem} \label{s31}

So far we have described the model in the weak coupling limit. We now
want to give a non-perturbative description of the background
for this orientifold. For this we focus our attention on the variation
of the dilaton-axion modulus field $\lambda$ as a function of the
coordinates $u$ and $v$. In the classical limit $\lambda\to i\infty$ near
the Dirichlet seven branes, and to $-i\infty$ near the orientifold
planes\cite{SENF}. This however cannot be the true in the full quantum 
theory since it follows from the definition \refb{e11b} of $\lambda$
that the imaginary part of $\lambda$ must be positive. Thus quantum effects
must modify $\lambda$. The question that we shall be asking is: what is
the fully quantum corrected $\lambda(u,v)$? This function must satisfy
the following two requirements:
\begin{itemize}
\item $Im(\lambda)$ must be positive everywhere in the complex $u,v$ plane.
\item $\lambda$ must be single valued in the $u,v$ plane up to an SL(2,Z)
transformation of the form:
\be \label{en3}
\lambda \to {p\lambda+q\over r\lambda+s}, \qquad \qquad p,q,r,s\in Z,
\qquad\qquad ps-qr=1.
\ee
\end{itemize}
This means that for every point in the complex $(u,v)$ plane we have
a torus whose modular parameter is given by $\lambda$. This torus is
described by an equation of the form:
\be \label{e12}
y^2 = x^3 + f(u,v) x + g(u,v)\, ,
\ee
with $\lambda$ being related to $f$ and $g$ through the relation:
\be \label{e12a}
j(\lambda) = {4 . (24f)^3\over 4f^3 + 27 g^2}\, .
\ee
$j(\lambda)$ is the modular function with a simple pole at
$\lambda=i\infty$.
Thus in order to determine $\lambda$ we need to determine the functions
$f(u,v)$ and $g(u,v)$. Note that $j(\lambda)$ blows up at the zeroes
of
\be \label{e12b}
\Delta(u,v) = 4 f^3 + 27 g^2 \, .
\ee
The locations of the zeroes of $\Delta$ can be identified as the
locations of seven branes in the full non-perturbative background.

One of the guiding principles in our attempt 
at determining $f$ and $g$ will be the fact that
as $u\to\infty$ at fixed $v$, the influence of the orientifold plane and the 
D-branes parallel to the $v$ plane, situated near $u=0$, must disappear
and the answer should reduce to the known answer for a configuration
of a single orientifold seven plane and four Dirichlet seven branes 
parallel to the $u$-plane\cite{SENF}. This gives for large $u$
\ben \label{e13}
f(u,v) &\simeq& \phi^2(u) f_{SW}(v; m'_1, m'_2, m'_1, m'_2, \tau) \cr
g(u,v) &\simeq& \phi^3(u) g_{SW}(v; m'_1, m'_2, m'_1, m'_2,\tau)
\een
where $\phi(u)$ is an arbitrary function of $u$ and $f_{SW}$ and
$g_{SW}$ are the functions that appear in describing the Seiberg-Witten
curve\cite{SEIWIT} for the N=2 supersymmetric SU(2) gauge theory with four 
hypermultiplets in the fundamental representation\cite{SENF}.
Similarly for large $v$  we must have
\ben \label{en4}
f(u,v) &\simeq& \wt\phi^2(v) f_{SW}(u; m_1, m_2, m_1, m_2, \tau) \cr
g(u,v) &\simeq& \wt \phi^3(v) g_{SW}(u; m_1, m_2, m_1, m_2,\tau)
\een
where $\wt\phi$ is an arbitrary function. Since $f_{SW}$ and $g_{SW}$ 
are polynomials of degree two and three respectively,
this motivates us to look
for polynomial solutions for $f$ and $g$, with $f$ quadratic in both 
$u$ and $v$ and $g$ cubic in both $u$ and $v$. In this case
the vacuum described by
eq.\refb{e12} appears to be very similar to an F-theory 
compactification of type IIB theory\cite{FTHEORY,VAFSIX}, 
but there are some differences which we shall point out later.

The problem of finding $f$ and $g$ also has an interpretation in terms of
non-perturbative dynamics on a three brane world-volume theory in the
spirit of refs.\cite{BDS,SEILAT,DOULAT}. For this
let us consider probing this configuration by a three brane lying in
the 0123 plane. The world-volume theory on the three brane 
has an $N=1$ space-time supersymmetry, with the three brane
coordinates $u$, $v$ and $x^4+ix^5$ serving as scalar components of
chiral superfields $U$, $V$ and $\Phi$ respectively. Of these fields
$\Phi$
decouples from the rest of the dynamics, so we shall focus on $U$ and
$V$. Besides these chiral superfields, for generic $u$ and
$v$ the three brane world-volume theory also has a U(1) gauge multiplet.
If $W_\alpha$ denotes the chiral superfield representing the gauge
field strength, then the low energy effective
field theory contains a gauge kinetic term of the form:
\be \label{e16}
\int d^4 x\int d^2\theta \, \lambda(U,V) W_\alpha W^\alpha\, + \, c.c.\, ,
\ee
where the function $\lambda(u,v)$ is the same function that we have
been trying to determine. Thus the problem that we are trying to address
can also be formulated as the problem of determining the effective gauge
coupling on the three brane world-volume theory.

The classical limit of this three brane world volume theory 
can be analysed along the lines of ref.\cite{COSH}.
For example, in the limit $u\to 0$,
the U(1) gauge symmetry on the three brane world volume theory gets
enhanced to SU(2), with the open string states streched between the
three brane and its image under $\II_{67}\cdot(-1)^{F_L}\cdot\Omega$
becoming massless. Let us denote this group by $SU(2)_1$.
On the other hand, for $v\to 0$ the same U(1)
gauge group is enhanced to another SU(2), whose massless charged gauge
bosons correspond to open string states stretched between the three
brane and its image under $\II_{89}\cdot(-1)^{F_L}\cdot\Omega$. This is
clearly a different SU(2) group, at least in this classical limit. 
Let us denote this by $SU(2)_2$.
In particular in the limit $u\to 0$,
$v\to 0$ we shall have the same U(1) group shared by two different SU(2)
groups! 
As shown in ref.\cite{COSH}, the unbroken gauge symmetry in this limit is
$SU(2)\times SU(2)$.
Later we shall see
how non-perturbative effects modify this picture.

\subsection{The Solution} \label{s32}

\subsubsection{$U(4)_v\times U(4)_u$ point} 
This point in the moduli space is the easiest to describe, since the
RR charge is neutralised locally, and hence we expect $\lambda(u,v)$
to be a constant independent of $u$ and $v$. {}From eq.\refb{e12a} we see
that this requires $f^3/g^2$ to be a constant. Since $f$ and $g$ are
assumed to be polynomials of degree 2 and 3 respectively in $u$ and $v$,
the most general solution is
\be \label{e14}
f(u,v)= \alpha u^2 v^2, \qquad g(u,v)= \beta u^3 v^3,
\ee
where we have used the freedom of shifting $u$ and $v$ by constants to
bring the zeroes of $f$ and $g$ to $u=v=0$. The constant value of
$\lambda$, which we shall denote by $\tau$, is given by the
equation:
\be \label{e15}
j(\tau)={4. (24\alpha)^3\over 4\alpha^3 + 27 \beta^2}\, .
\ee
One of the constants $\alpha$ and $\beta$ can be absorbed in a rescaling 
of the form 
\be \label{e26}
f\to k^2 f, \qquad \qquad g\to k^3 g\, ,
\ee
for any constant $k$.  This leaves the ratio $f^3/g^2$
fixed. Thus this vacuum is characterised by only one complex
parameter $\tau$ as expected.

Note that the background described by \refb{e14} looks very similar
to an F-theory background with a pair of intersecting $D_4$ 
singularities. However there are some differences. First of all, as we
move around the origin in  the $u$ ($v$) plane, there is an $SO(8)_u$
($SO(8)_v$) monodromy given by the $SO(8)$ matrix $\MM$ defined in \refb{e1}.
This breaks the gauge group to $U(4)_u$ ($U(4)_v$). In the language of
ref.\cite{VAFSIX,OTHERS} this would correspond to an inner automorphism 
of $SO(8)$,
which is normally taken to be absent in conventional F-theory background.
Furthermore, in the conventional F-theory vacuum, an intersection of
two $D_4$ singularities produces a collapsed two cycle, and hence a 
tensionless string associated with a three brane wrapped around the
two cycle\cite{VAFSIX}. 
In the present case the collapsed two cycle is associated with the
$Z_2$ orbifold singularity obtained by modding out $R^4$ by the 
group element $gh$. However, as was shown by Aspinwall\cite{ASPINK3},
in the conformal field theory orbifold  we have half unit of the
$B_{\mu\nu}$ flux through the collapsed two cycle, so that the Kahler class
associated with the two cycle, instead of vanishing, is purely imaginary.
For type IIA theory,
this prevents the masses of two branes wrapped on the two cycle from
vanishing. By T-duality we expact the same mechanism to prevent the
tension of the three brane wrapped on the two cycle to vanish for 
conformal field theory orbifolds.

\subsubsection{$SU(2)_v\times SU(2)'_v\times U(4)_u$ Gauge Group}

In the classical limit this point is obtained by pulling the four seven
branes parallel to the $v$ plane away from the orientifold plane at $u=0$,
keeping the seven brane positions parallel to the $u$ plane intact. 
(Here we are following the convention introduced in the previous section,
according to which the subscript $u$ ($v$) labels the gauge fields living
on the branes parallel to the $u$ ($v$) plane, {\it i.e.} transverse to
the $v$ ($u$) plane.) Since
the RR charge associated with branes transverse to the $v$ plane is still
locally neutralised, we expect $\lambda$ to be independent of $v$. Thus
the $v$ dependence of $f$ and $g$ should still be of the form $v^2$ and 
$v^3$ respectively. The $u$ dependence of $f$ and $g$ can then be 
determined by going to the large $v$ limit, in which case the $f$ and $g$
must satisfy the boundary condition \refb{en4}.
This gives, for all $v$, 
\ben \label{e17}
f(u,v) &=& v^2 f_{SW}(u; m_1, m_2, m_1, m_2, \tau)\cr
g(u,v) &=& v^3 g_{SW}(u; m_1, m_2, m_1, m_2, \tau)
\een
{}From eq.\refb{e12b} we get
\be \label{e19}
\Delta(u,v) = v^6(4 f_{SW}(u)^3 + 27 g_{SW}(u)^2)\equiv 
v^6 \Delta_{SW}(u; m_1, m_2, m_1, m_2, \tau)\, .
\ee
{}From the analysis of ref.\cite{SEIWIT} we know that for $m_1=m_3$
and $m_2=m_4$, $\Delta_{SW}$ has two second order zeroes and two first
order zeroes, without $f$ and $g$ vanishing at those points. Locally
this would correspond to a $U(2)_v\times U(2)'_v$ non-abelian
gauge symmetry living on the branes parallel to the $v$ plane. (In the
orientifold description, these
would be generated by the $SO(8)$ generators commuting with the Higgs
vev given by eqs.\refb{e5}, \refb{e10}.)
The $SO(8)_v$ monodromy \refb{e1} around the origin in the $v$ plane  
breaks this to $SU(2)_v\times SU(2)'_v$. Note that since in
the $v$-plane all the zeroes of
$\Delta$ are at $v=0$, there is no further monodromy.
For $m_1=m_2$, $\Delta_{SW}$ has a fourth order zero and two
first order zeroes, signalling a local $U(4)_v$ gauge group. Again this
is broken to $Sp(4)_v$ by the $SO(8)_v$ monodromy \refb{e1}.

The $D_4$ singularity at $v=0$, signalled by the zeroes of order 
two, three and six in $f$, $g$ and $\Delta$ respectively, implies that
locally there is an $SO(8)_u$ gauge symmetry living on the $v=0$ plane.
This breaks to $U(4)_u$ by the $SO(8)_u$ monodromy \refb{e1} at $u=\infty$. 
However, since there are now several singularities in the $u$ plane, we need
to ensure that monodromy around these singularities does not break the 
$U(4)_u$ gauge group any further. Otherwise there will be a discrepancy 
between the perturbative and non-perturbative description of the model
signalling that the non-perturbative description that we are proposing is
not correct.
In particular, since now there is non-trivial SL(2,Z) monodromy in the
$u$ plane, these will induce triality automorphisms 
in $SO(8)_u$\cite{SEIWIT,SENF}, and can break $U(4)_u$ further unless 
these automorphisms
commute with $U(4)_u$\cite{VAFSIX}. We shall now show that this does not
happen. Since the $U(1)$ factor of $U(4)_u$ is broken in any case by
anomaly effects, we shall focus on the $SU(4)_u$ subgroup of $U(4)_u$.

To analyse these monodromies, let us
consider the generators of SL(2,Z) transformation:
\be \label{e20}
S = \pmatrix{0 & 1 \cr -1 & 0},  \qquad \qquad T=\pmatrix{1 & 1\cr 0 & 1}\, .
\ee
As was shown in \cite{SEIWIT}, $S$ and $T$ induce the following
triality transformations on the three $SO(8)$ representations $8_v$, $8_s$
and $8_c$:
\ben \label{en5}
T &:& 8_s\leftrightarrow 8_c, \qquad 8_v\to 8_v\cr
S &:& 8_s\leftrightarrow 8_v, \qquad 8_c\to 8_c
\een
Around the double zeroes of $\Delta$ in the $u$-plane
we get an SL(2,Z) monodromy 
conjugate to $T^2$, which, from \refb{en5}, can be seen to have
trivial $SO(8)$ monodromy.\footnote{This reflects a consistency check
for the model. If the orientifold model allowed a deformation that
separates all the four seven branes parallel to the $v$ plane away from
each other, then it would be impossible to keep $SU(4)_u$ unbroken
under this deformation since there will be non-trivial $SU(4)_u$
monodromy around these seven branes.} Thus we need to focus our attention
on the monodromy around the single zeroes of $\Delta_{SW}$ in the $u$
plane. Since the product of these two monodromies must equal \refb{e1}
which has been already taken into account, we only need to focus on the
monodromy around one of the single zeroes of $\Delta$. Let us focus on the
singularity that corresponds to a massless monopole\cite{SEIWIT}. The
SL(2,Z) monodromy around this point is given by $STS^{-1}$. {}From
eq.\refb{en5} we see that this corresponds to SO(8) automorphism:
\be \label{en6}
8_v\leftrightarrow 8_c, \qquad 8_s\to 8_s\, .
\ee
Let us now study the decomposition of these representations under 
$SU(4)_u\subset SO(8)_u$, embedded as described in eqs.\refb{e1}, \refb{e2}. 
It is as follows:
\ben \label{e21}
8_v & \to & 4 + \bar 4 \cr
8_c & \to & 4 + \bar 4\cr
8_s & \to & 6 + 2\, .
\een 
Thus we see that the monodromy around the single zero of $\Delta$ commutes
with the $SU(4)_u$ group and does not break it any further.

Another way of interpreting this result is as follows. The monodromy 
\refb{en6} breaks $SO(8)_u$ to $SO(7)_u$\cite{VAFSIX}. Thus the final
unbroken gauge group should be the intersection of $SO(7)_u$ and
$SU(4)_u$. What the above analysis shows is that it is consistent to
embedd the $SO(7)_u$ in $SO(8)_u$ in such a way that it contains the
$SU(4)_u\equiv SO(6)_u$ subgroup of $SO(8)_u$.

\subsubsection{$Sp(4)_v\times Sp(4)_u$ Gauge Group}

In the classical limit this corresponds to pulling the four seven branes
parallel to the $u$ plane away from the orientifold plane $v=0$ keeping
them together, and at the same time pulling the four seven branes
parallel to the $v$ plane away from the orientifold plane $u=0$ keeping
them together. Locally there is a $U(4)_u$ ($U(4)_v$) gauge group living 
on the seven branes parallel to the $u$ ($v$) plane which is broken to
$Sp(4)_u$ ($Sp(4)_v$) by the monodromy \refb{e1} in the $u$ ($v$) plane
at $u=\infty$ ($v=\infty$). In order to get these gauge groups in the
non-perturbative description
we must choose $f$ and $g$ in such a way that there is an $A_3$ singularity
parallel to the $u$ plane and an $A_3$ singularity parallel to the $v$
plane. This corresponds to a fourth order zero of $\Delta$ at $u=u_0$ and
a fourth order zero
at $v=v_0$ for some $u_0$, $v_0$. Thus we need a $\Delta$ of the
form:
\be \label{e22}
\Delta(u,v) = (u-u_0)^4 (v-v_0)^4 \delta(u,v)\, ,
\ee
where $\delta$ is a polynomial of degree two in $u$ and $v$. Thus the 
question now is: is it possible to choose $f$ and $g$ so that $\Delta$
defined in eq.\refb{e12b} has the form \refb{e22}?

At the first sight it would seem unlikely that such $f$ and
$g$ exist. $\Delta$ is a polynomial of degree six in both $u$
and $v$. A generic polynomial of this kind is labelled by 49 parameters.
On the other hand the right hand side of \refb{e22} is labelled by
11 parameters, two from $u_0$ and $v_0$, and 9 from $\delta(u,v)$. 
Thus if we start with a generic $f$ and $g$, requiring $\Delta$ to be of 
the form \refb{e22}
would give $49-11=38$ constraints on the coefficients appearing
in $f$ and $g$. Since the total number of parameters appearing in $f$
and $g$ is $9+16=25$, we get a set of 38 equations for 25 parameters.
This is a hightly overdetermined system of equations!

Nevertheless it turns out that there does exist a family of 
solutions to this
system of equations. As has already been stated earlier, the main guide 
for obtaining these solutions is the use of boundary conditions
\refb{e13}, \refb{en4} with $m_1=m_2$, $m_1'=m_2'$. 
The relevant $f_{SW}$ and $g_{SW}$ 
appearing in these boundary conditions are given by
\ben \label{e29}
f_{SW}(u; m,m,m,m,\tau) & = & c^2 m^4 \wt f_{SW}
\Big({u\over m^2 c}-{b-3\over 2}; \tau\Big) \cr\cr
g_{SW}(u; m,m,m,m,\tau) & = & c^3 m^6 \wt g_{SW}
\Big({u\over m^2 c}-{b-3\over 2}; \tau\Big) \, ,
\een
where 
\be \label{e22b}
b = -{3\vartheta^4_2(\tau)\over \vartheta^4_1(\tau)+\vartheta^4_3(\tau)},
\qquad \qquad c = -{1\over 3} 
(\vartheta^4_1(\tau)+\vartheta^4_3(\tau))\, ,
\ee
\ben \label{e30}
\wt f_{SW}(\wt u; \tau) & = & -{1\over 3} b^4 + 2 b^2 \wt u - {1\over 4}
(3+b^2) \wt u^2 \cr\cr
\wt g_{SW}(\wt u; \tau) & = & {1\over 108} (-b^2 + 3\wt u) (8 b^4
-48 b^2 \wt u + 9(b^2 -1) \wt u^2)\, .
\een
$\vartheta_i$ are the Jacobi theta functions.
In extracting $f_{SW}$ and $g_{SW}$ from ref.\cite{SEIWIT} we have used the
rescaling freedom \refb{e26}. 

The general solution for $f(u,v)$ and $g(u,v)$ (up to 
rescaling of $f$ and $g$ of the form\refb{e26})
satisfying these boundary
conditions and giving a $\Delta$ of the form \refb{e22} is 
\ben \label{e23}
f(u,v) &=& m^4 (m')^4 c^4 \wt f\Big({u\over m^2 c}-{b-3\over 2}\, , \,
{v\over (m')^2 c}-{b-3\over 2}\, ; \tau\Big) \cr \cr
g(u,v) &=& m^6 (m')^6 c^6 \wt g\Big({u\over m^2 c}-{b-3\over 2}\, , \,
{v\over (m')^2 c}-{b-3\over 2}\, ; \tau\Big) \, ,
\een
where
\ben \label{e23a}
\wt f(\wt u, \wt v; \tau) & = & {1\over 12}\,  
[-\alpha^2 - 4\alpha b^2 (\wt u + 
\wt v) - 4 b^4 (\wt u^2 + \wt v^2) + (12\alpha - 8b^4) \wt u \wt v\nonumber \\
&& + 24 b^2 \wt u \wt v (\wt u + \wt v) -(9+3b^2) \wt u^2 \wt v^2]\cr 
\nonumber \\
\wt g(\wt u, \wt v; \tau) &=& {1\over 216} \, [-\alpha - 2b^2(\wt u + \wt v)
+ 6\wt u \wt v] \nonumber \\
&& \times [2\alpha^2 + 8\alpha b^2(\wt u + \wt v) + 8b^4(\wt u^2 +
\wt v^2) + (16 b^4 - 24 \alpha) \wt u \wt v \nonumber \\
&& - 48 b^2 \wt u \wt v 
(\wt u+\wt v) + 9(b^2 -1) \wt u^2 \wt v^2]\, .
\een
Here $\alpha$ is an arbitrary complex parameter whose significance will 
be explained later.

It is easy to verify that $f$ and $g$ defined this way satisfy the
boundary conditions \refb{e13}, \refb{en4} with $m_1=m_2$ and
$m_1'=m_2'$. Also $\Delta$ computed from $f$ and $g$ given in eq.\refb{e23}
is given by
\ben \label{e25}
\Delta(u,v) & = & - {1\over 16} b^2 (9-b^2)^2 
\Big( u -{1\over 2}(b-3) \, c \, m^2\Big)^4
\Big( v -{1\over 2}(b-3) \, c\, (m')^2\Big)^4 \cr
&& \Big[ uv - c\Big( {b+3\over 2} +\sqrt{9-b^2}\Big) (u(m')^2 + v m^2)
\cr
&& + \Big\{ {1\over 4} (b-3)^2 + (3 + \sqrt{9-b^2}) (b-3) 
-{\alpha\over 2b^2}
(3 + \sqrt{9-b^2}\Big\} m^2 (m')^2 c^2\Big]\cr
&& \Big[ uv - c\Big( {b+3\over 2} -\sqrt{9-b^2}\Big) (u(m')^2 + v m^2)
\cr
&& + \Big\{ {1\over 4} (b-3)^2 + (3 - \sqrt{9-b^2}) (b-3) 
-{\alpha\over 2b^2}
(3 - \sqrt{9-b^2}\Big\} m^2 (m')^2 c^2\Big]\cr
&& \een
This shows that $\Delta$ does indeed have the form of \refb{e22} with $u_0$
and $v_0$ given by $(b-3)cm^2/2$ and $(b-3)c(m')^2/2$ respectively.

Let us first examine the existence of $Sp(4)_v\times Sp(4)_u$ gauge group
by examining the singularity structure of this configuration. In terms
of coordinates $\tu$, $\tv$ defines as
\be \label{eq1}
\tu = {u\over cm^2} - {1\over 2}(b-3), \qquad \qquad \tv = 
{v\over c(m')^2} -{1\over 2} (b-3)\, ,
\ee
$\Delta$ has a fourth order zero at $\tv=0$. $f$ and $g$ can be
rewritten in this coordinate system as
\ben \label{eq2}
f(u,v) &=& -3 m^4 (m')^4 c^4 (h_1^2(\tu) + h_1(\tu) h_2(\tu) \tv
+ h_3(\tu) \tv^2)\cr\cr
g(u,v) &=& m^6 (m')^6 c^6 \Big( 2 h_1^3(\tu) + 3 h_1^2(\tu) h_2(\tu)
\tv +3 \big( h_3(\tu) + {1\over 4} h_2^2(\tu)\big) h_1(\tu) \tv^2 \cr
&& + \big( {3\over 2} h_3(\tu) - {1\over 8} h_2^2(\tu)\big) h_2(\tu)
\tv^3\Big)\, ,
\een
where,
\ben \label{eq3}
h_1(\tu) &=& -{1\over 6} (\alpha + 2 b^2 \tu) \cr \cr
h_2(\tu) &=& {2\over 3} (-b^2 + 3\tu) \cr \cr
h_3(\tu) &=& {1\over 9} b^4 -{2\over 3} b^2 \tu + {1\over 12} 
(b^2 +3)\tu^2\, .
\een
Comparing with the results of ref.\cite{VAFSIX} we see that 
this corresponds to a non-split $A_3$ singularity. (A split $A_3$
singularity will require $h_1$ to have only double zeroes in the
$\wt u$ plane.) Thus the gauge group living on the $\tv =0$ plane is 
$Sp(4)_u$ as expected. An identical analysis shows that the gauge group
living on the $\tu=0$ plane is also $Sp(4)_v$ as expected.

Next we discuss the interpretation of the complex parameter $\alpha$.
{}From eqs.\refb{e23}, \refb{e23a} it is clear that in the limit $m\to 0$
or $m'\to 0$ this parameter disappears from the expressions for $f(u,v)$
and $g(u,v)$. In other words if either $U(4)_u$ or $U(4)_v$ is
unbroken at the classical level, 
then we do not have the deformation of the non-perturbative
background labelled by
$\alpha$, whereas if both $U(4)$'s are broken then this deformation of
the background is present. This is precisely what we expect for the
deformation associated with the massless closed string state from the
twisted sector. When either of the $U(1)$ factors is present in the
classical theory, then due to one loop anomaly this $U(1)$ gauge
field becomes massive by absorbing these twisted sector closed 
string states;
whereas if both $U(1)$'s are broken by Higgs mechanism at the classical
level, then these twisted sector closed string states remain massless
and act as moduli field\cite{WITSIX}. 
Furthermore, from eq.\refb{e23a} we see that the deformation associated
with $\alpha$ does not affect the form of $f$ and $g$ for large $u$ or
large $v$, again as is expected of a blow up mode localised near the
orbifold point.
Thus it is very likely that the parameter
$\alpha$ is related to the deformations associated with the twisted
sector closed string states (the blow up modes of the orbifold
singularity) although we do not have a direct proof of this statement.

The geometry of seven brane configurations
described by this non-perturbative background can be studied by examining
the zeroes of $\Delta$. {}From eq.\refb{e25} we see that there
are four coincident seven branes at $u=(b-3)m^2c/2$ and four
coincident seven branes at $v=(b-3)(m')^2c/2$. This is identical to the
configuration of seven branes found in the classical limit. Also,
in the $\tau\to i\infty$ limit, 
\be \label{e37}
b\to (-3)\, , \qquad \qquad c\to -(1/3)\, ,
\ee
as can be seen from \refb{e22b}. Thus 
the locations of these coincident seven
branes approach $u\simeq m^2$ and $v\simeq (m')^2$, exactly as expected
from the orientifold description.

More interesting is the configuration of seven branes coming from the
second factor of $\Delta$. {}From eq.\refb{e25} we see that we have two
more seven branes situated on the surfaces:
\ben \label{e34}
&& \Big[ uv - c\Big( {b+3\over 2} \pm\sqrt{9-b^2}\Big) (u(m')^2 + v m^2)
+ \Big\{ {1\over 4} (b-3)^2 + (3 \pm \sqrt{9-b^2}) (b-3) \cr
&& -{\alpha\over 2b^2}
(3 \pm \sqrt{9-b^2}\Big\} m^2 (m')^2 c^2\Big] = 0\, .
\een
For large $v$, this equation reduces to 
\be \label{e35}
u\simeq \Big( {b+3\over 2} \pm \sqrt{9-b^2}\Big) cm^2\, .
\ee
These two surfaces simply represent the two seven branes into which an
orientifold plane parallel to the $v$ plane would split in the absence
of the projection $(-1)^{F_L}\cdot\Omega\cdot\II_{89}$. Similarly for 
large $u$ eq.\refb{e34} becomes
\be \label{e36}
v\simeq \Big( {b+3\over 2} \pm \sqrt{9-b^2}\Big) c(m')^2\, .
\ee
These two surfaces represent the two seven branes into which an
orientifold plane parallel to the $u$ plane would split in the absence
of the projection $(-1)^{F_L}\cdot\Omega\cdot\II_{67}$. 

The phenomenon of the orientifold plane splitting into two seven
branes under non-perturbative quantum corrections is not 
new\cite{SENF}. What is
new is the phenomenon that the two seven branes into which the
orientifold plane parallel to the $u$ plane splits smoothly join the
two seven branes into which the orientifold plane parallel to the
$v$ plane splits. Thus at the end we get only two seven branes instead
of two pairs of intersecting seven branes.\footnote{Of 
course at special values
of $\alpha$ given by
$\alpha = -2 b^2 (3 \pm \sqrt{9-b^2})$
one of these two seven branes degenerates into two intersecting 
seven branes.}

These two seven branes intersect at the points
\ben \label{es1}
\Big( u = {b-3\over 2} cm^2,\, \,
v = \Big( {b-3\over 2} -{\alpha\over 2b^2}\Big)
c (m')^2\Big)\, , \nonumber \\ 
\Big( u = \Big( {b-3\over 2} -{\alpha\over 2b^2}\Big)
c m^2,\, \, v = {b-3\over 2} (cm')^2 \Big)\, .
\een
The physical interpretation of these intersection points is as
follows. Since at $u=(b-3) cm^2/2$ there are four coincident seven branes,
$\lambda(u,v)\to i\infty$ on this plane. Thus the phenomenon of the splitting
of the orientifold plane at $v=0$ must disappear for $u=(b-3) cm^2/2$.
and the two seven branes into which the orientifold plane splits
must meet on this plane. Similarly the two seven branes into which the
$u=0$ orientifold plane splits must meet at $v=(b-3)c (m')^2/2$.

We can now use these results to extract information about the
non-perturbative dynamics on the corresponding three brane world-volume
theory. We shall concentrate on the infrared dynamics on the three brane
world volume theory when it is close to one of the two seven branes given
in eq.\refb{e34}.
In this case
the corresponding infrared dynamics is governed by an N=2 supersymmetric
theory on the world volume with a U(1) vector multiplet and a charged
hypermultiplet representing open string states stretched between the
three brane and the seven brane\cite{SENBPS}. 
If we move the three brane along
such a seven brane to $u\to \infty$, then the massless charged hypermultiplet
that we get can be interpreted as the monopole / dyon state 
associated with the $SU(2)_2$ gauge group on the three brane world volume
introduced at the end of subsection \ref{s31}.
On the other hand we can move the three
brane along the same seven brane to $v\to\infty$ without hitting any
singularity, and in this case the same massless hypermultiplet can be
interpreted as the monopole / dyon associated with the $SU(2)_1$ gauge group
introduced at the end of subsection \ref{s31}. Thus we see that
once non-perturbative effects are taken into account, an $SU(2)_1$
monopole can be continuously transformed into an $SU(2)_2$ monopole and
vice versa, although classically they correspond to two distinct gauge
groups.

In the weak coupling limit \refb{e37} 
both the seven branes given in \refb{e34} coincide and are described by
the equation:
\be \label{e38}
uv -\Big(1+{1\over 54}\alpha\Big) m^2 (m')^2 =0\, .
\ee
This reflects the fact the phenomenon of the splitting of the orientifold
plane disappears in the classical limit. However for generic value of
$\alpha$, \refb{e38}
still describes a smooth surface instead of a pair of intersecting
orientifold planes as in ref.\cite{GIMPOL}.  Only for $\alpha=-54$ 
\refb{e38} takes the form
\be \label{en7}
uv = 0,
\ee
describing a pair of intersecting orientifold planes. Thus the
orbifold limit in weak coupling must correspond to $\alpha=-54$.
A generic $\alpha$ corresponds to a blown up version of
the orbifold singularity, and hence we
do not expect a singular intersection at $u=v=0$.  This reconfirms
our interpretation of $\alpha$ as the vev of twisted sector closed
string states.

\subsubsection{$SU(2)_v\times SU(2)'_v\times Sp(4)_u$ Gauge Group}

In the classical limit, this corresponds to pulling apart the four
seven branes parallel to the $v$ plane into two pairs of seven branes
situated at $u = m_1^2$ and $u=m_2^2$. Since enhanced SU(2) gauge symmetry
requires $\Delta$ to have zeroes of order two, we are looking for $f(u,v)$
and $g(u,v)$ such that $\Delta$ defined in eq.\refb{e12b} takes the form
\be \label{er1}
\Delta(u,v) = (u-u_1)^2 (u-u_2)^2 (v-v_0)^4 \widehat \delta(u,v)\, ,
\ee
where $u_1$, $u_2$ and $v_0$ are constants, and $\widehat\delta$ is a
polynomial of degree (2,2) in $u$ and $v$. The most general $f$ and $g$
satisfying \refb{er1} and the required asymptotic behavior at infinity
is given by
\ben \label{er2}
f(u,v) &=& m_1^2 m_2^2 (m')^4 c^4 \wt f(\tu, \tv) \cr
g(u,v) &=& m_1^3 m_2^3 (m')^6 c^6 \wt g(\tu, \tv) \, , 
\een
where $b$, $c$ are as defined in eq.\refb{e22b}, and,
\be \label{er3a}
\tu = {u\over m_1 m_2 c} - {1\over 2} (b-3)\eta, \qquad \qquad
\tv = {v\over (m')^2 c} - {1\over 2} (b-3)\, ,
\ee
\be \label{er3}
\eta = {1\over 2} \Big( {m_1\over m_2} + {m_2\over m_1}\Big)\, ,
\ee
\ben \label{er4}
\wt f(\tu,\tv) &=& -3 (h_1^2(\tu) + h_1(\tu) h_2(\tu) \tv
+ h_3(\tu) \tv^2)\cr\cr
\wt g(\tu,\tv) &=& \Big( 2 h_1^3(\tu) + 3 h_1^2(\tu) h_2(\tu)
\tv +3 \big( h_3(\tu) + {1\over 4} h_2^2(\tu)\big) h_1(\tu) \tv^2 \cr
&& + \big( {3\over 2} h_3(\tu) - {1\over 8} h_2^2(\tu)\big) h_2(\tu)
\tv^3\Big)\, ,
\een
where,
\ben \label{er5}
h_1(\tu) &=& -{1\over 6} (\alpha + 2 b^2 \tu) \cr
h_2(\tu) &=& {2\over 3} (-b^2\eta + 3\tu) \cr
h_3(\tu) &=& {1\over 36} b^4 (3 + \eta^2) -{3\over 4} b^2(1 - \eta^2)
-{2\over 3} b^2 \eta \tu + {1\over 12} (b^2 +3)\tu^2\, .
\een
$\Delta(u,v)$ calculated from $f$ and $g$ given in eqs.\refb{er2}-\refb{er5}
is given by
\ben \label{er6}
\Delta(u,v) &=& -{1\over 16} b^2 (9-b^2)^2 m_1^6 m_2^6 (m')^{12} c^{12}
(\wt u^2 + b^2 
(1-\eta^2))^2 \tv^4 \nonumber \\
&& \Big[ \tu^2 \tv^2 - 6\tu\tv (\tu + \eta\tv) + b^2 (\tu^2 + \tv^2)
+ 9(\eta^2-1) \tv^2\nonumber \\
&& + (2 b^2 \eta - {3\alpha\over b^2}) \tu\tv + \alpha (\tu + \eta \tv) +
{\alpha^2\over 4b^2}\Big]\, .
\een
{}From this we see that $\Delta$ has two second order zeroes at
\be \label{er7}
u = \Big( {b-3\over 2}\eta \pm b \sqrt{\eta^2-1}\Big) m_1 m_2 c\, ,
\ee
signalling the presence of $SU(2)_v\times SU(2)_v'$ gauge symmetry group.
It also has a fourth order zero at
\be \label{er8}
v = {b-3\over 2} (m')^2 c\, .
\ee
{}From the structure of $f(u,v)$ and $g(u,v)$ given in 
eqs.\refb{er2}-\refb{er5} we see that this corresponds to a non-split
$A_3$ singularity and hence $Sp(4)_u$ gauge group.

Finally we note that the last factor of $\Delta$ given 
in eq.\refb{er6} does not factorise into two factors as in eq.\refb{e25}.
This shows that the pair of seven branes into which the intersecting pair
of orientifold planes split now further join together smoothly to give one
single seven brane.

\subsubsection{$SU(2)_v\times SU(2)'_v\times SU(2)_u\times SU(2)'_u$ 
Gauge Group}

We could further split the
four seven branes parallel to the $u$ plane into two pairs to
obtain the $SU(2)_v\times SU(2)'_v\times SU(2)_u\times SU(2)'_u$ model.
This model will be characterised by four complex parameters $m_1$, $m_2$,
$m_1'$ and $m_2'$ labelling the positions of the four seven brane pairs
in the classical limit $-$ two parallel to the $v$-plane and two
parallel to the $u$-plane $-$ besides the parameters $\tau$ and $\alpha$.
We now need to look for $f$ and $g$ such that $\Delta$
defined in eq.\refb{e12b} takes the form:
\be \label{en8}
\Delta(u,v) = (u-u_1)^2 (u-u_2)^2 (v-v_1)^2 (v-v_2)^2 \wt \delta(u,v)\, ,
\ee
where $u_i$ and $v_i$ are arbitrary constants and $\wt\delta$ is a
polynomial of degree two in $u$ and $v$. 
A family of solutions for $f$ and $g$ satisfying this criteria is given by
\ben\label{ez0}
f(u,v) &=& m_1^2 m_2^2 (m_1')^2 (m_2')^2 c^4 \tf(\tu, \tv)\cr
g(u,v) &=& m_1^3 m_2^3 (m_1')^3 (m_2')^3 c^6 \tg(\tu, \tv)\, ,
\een
where,
\ben \label{ez1} 
\tf(\tu,\tv) &=& {1\over 12} 
(-\alpha^2 + 27 b^4 ((\eta')^2-1) (\eta^2-1) 
- 3 b^6 ((\eta')^2-1) (\eta^2-1) - 4 \alpha b^2 \eta' \tu \cr && 
+ 27 b^2 (1 - (\eta')^2) \tu^2 - b^4 (3 + (\eta')^2) \tu^2 - 
4 \alpha b^2 \eta \tv 
+ 12 \alpha \tu \tv - 8 b^4 \eta' \eta \tu \tv \cr &&
+ 24 b^2 \eta' \tu^2 \tv 
+ 27 b^2 (1-\eta^2) \tv^2 - b^4 (3+\eta^2) \tv^2 + 
    24 b^2 \eta \tu \tv^2 - 9 \tu^2 \tv^2 - 3 b^2 \tu^2 \tv^2) \cr\cr
\tg(\tu,\tv) &=& {1\over 216}
     (-\alpha - 2 b^2 \eta' \tu - 2 b^2 \eta \tv + 6 \tu \tv) \cr &&
     [2 \alpha^2 - 81 b^4 ((\eta')^2-1) (\eta^2-1) + 
      9 b^6 ((\eta')^2-1) (\eta^2-1) + 8 \alpha b^2 \eta' \tu \cr &&
+ b^4 (9 - (\eta')^2) \tu^2 + 
      81 b^2 ((\eta')^2-1) \tu^2 + 8 \alpha b^2 \eta \tv 
- 24 \alpha \tu \tv + 16 b^4 \eta' \eta \tu \tv \cr &&
- 48 b^2 \eta' \tu^2 \tv + 81 b^2 (\eta^2-1) \tv^2 + b^4 (9-\eta^2) \tv^2 
- 48 b^2 \eta \tu \tv^2 
+ 9 (b^2-1) \tu^2 \tv^2]\, , \cr  &&
\een
\be \label{ez2}
\eta = {1\over 2} \Big( {m_1\over m_2} + {m_2\over m_1}\Big)\, , 
\qquad \qquad
\eta' = {1\over 2} \Big( {m'_1\over m'_2} + {m'_2\over m'_1}\Big)\, , 
\ee
and,
\be \label{ez3}
\tu = {u\over m_1 m_2 c} - {1\over 2} (b-3)\eta, \qquad \qquad
\tv = {v\over m'_1 m'_2 c} - {1\over 2} (b-3) \eta'\, .
\ee
$\Delta$ computed from this $f$ and $g$ is given by,
\ben \label{ez5}
\Delta &=& {1\over 64} m_1^6 m_2^6 (m_1')^6 (m_2')^6 c^{12}
(-9 + b^2)^2 (b^2 - b^2 \eta^2 + \tu^2)^2 
(b^2 - b^2 (\eta')^2 + \tv^2)^2  \cr &&
    \big[-\alpha^2 +  4b^4 (9-b^2) (1-\eta^2) (1-(\eta')^2)
- 4 \alpha b^2 \eta' \tu 
\cr &&
+ 
36 b^2 (1 - (\eta')^2)\tu^2 - 4 b^4 \tu^2 - 4 \alpha b^2 \eta \tv 
+ 12 \alpha \tu \tv -
      8 b^4 \eta' \eta \tu \tv + 24 b^2 \eta' \tu^2 \tv \cr &&
+ 36 b^2 (1 - \eta^2) \tv^2 - 4 b^4 \tv^2  
      + 24 b^2 \eta \tu \tv^2 - 4 b^2 \tu^2 \tv^2\big]\, .
\een

The geometry of the seven brane configurations representing the split
orientifold plane does not have any new
feature that was not already present in the previous examples. Note that
in this expression if we take one of the $m_i$'s or one of the $m_i'$'s
to zero, the $\alpha$ dependence drops out from $f$ and $g$. In this
case, in the classical limit we recover an $U(2)$ gauge 
group\cite{GIMPOL}. The presence of the $U(1)$ factor implies that the
twisted sector closed string states would become massive due to anomaly
effects. This is again consistent with the interpretation of $\alpha$ as
the blow up mode.

\subsubsection{More General Deformations}

We can further deform the model by switching on the vev of the 
hypermultiplets in the (4,4) representation of $U(4)_v\times U(4)_u$.
Since these hypermultiplets are localised at the intersection of the
seven branes in the classical limit, we expect that the vev of these
hypermultiplets should not change the asymptotic form of $f$ and $g$
for large $u$ or large $v$. Thus switching on vev of these
hypermultiplets should correspond to addition of terms in $f$ and $g$
of the form:
\ben \label{ep1}
\delta f(u, v) &=& \sum_{m,n=0}^1 a_{mn} u^m v^n\, ,\cr
\delta g(u, v) &=& \sum_{m,n=0}^2 b_{mn} u^m v^n\, ,
\een
This will generically break the gauge group completely, and hence must
describe a configuration where the seven branes, which previously
existed only in pairs, are separated from each other. Naively it might
sound like a violation of the condition \refb{e10} and hence of
\refb{e7}, but this is not so. To see how this contradiction is
avoided, let us note that given any $f(u,v)$ and $g(u,v)$ which are
of degree (2,2) and (3,3) respectively, we can express them as
\ben \label{en9}
f(u,v) & = & f_{SW}(u; m_1(v), m_2(v), m_3(v), m_4(v), \tau(v))\, ,\cr
g(u,v) & = & g_{SW}(u; m_1(v), m_2(v), m_3(v), m_4(v), \tau(v))\, ,
\een
after suitable $v$ dependent shift of $u$, and 
$v$ dependent rescaling of $f$ and $g$ that keeps 
$\lambda(u,v)$ invariant. Here $m_i$ are in general functions of $v$.
Eq.\refb{e7} then implies that
\be \label{en10}
m_1(z) = m_3(-z), \qquad \qquad m_2(z)=m_4(-z)\, ,
\ee
where $z=\sqrt{v}$.
Thus there is no need for $m_1$ and $m_3$ (or $m_2$ and $m_4$) to be equal
as long as they are allowed to vary with $v$.

To see how eq.\refb{en9} is realised in practice, let us consider perturbing
the $SU(2)_v\times SU(2)_v'\times SU(2)_u\times SU(2)_u'$ model by
adding terms of the form \refb{ep1} to $f$ and $g$. This will, in general,
split the double zeroes of $\Delta$ given in \refb{en8}. Let us focus on the
zeroes of $\Delta$ near $u=u_1$. After addition of \refb{ep1} to $f$ and $g$,
$\Delta$ near $u=u_1$ for large $v$ takes the form:
\be \label{ep2}
\Delta\simeq C v^6 (u-u_1)^2 - \epsilon v^5\, .
\ee
Here $C$ and $\epsilon$ are constants, with $\epsilon$ associated with
the hypermultiplet vev. The crucial point is that due to the nature of
the form \refb{ep1} of $\delta f$ and $\delta g$, the coefficient of
$\epsilon$ for large $v$ is of order $v^5$ and not of order $v^6$.
The zeroes of $\Delta$ now get shifted to
\be \label{ep3}
u \simeq u_1 \pm \sqrt{\epsilon/Cv}\, ,
\ee
for large $v$. In the weak coupling limit these may be identified to 
$m_1^2(v)$ and $m_3^2(v)$ respectively. Thus we see that under $v\to
e^{2\pi i} v$, $m_1^2(v)$ and $m_3^2(v)$ get interchanged as required.
Similar reasoning shows that other $m_i^2$ and $(m'_i)^2$'s also satisfy the
required monodromy in the $u$ / $v$ plane.

One can also consider special subspaces of the full moduli space where
some of the diagonal $SU(2)$ subgroups are unbroken. Consider for example
the hypermultiplet transforming in the $(2,2)$ representation of 
$SU(2)_u\times SU(2)_v$. In this case we can switch on the vev of the 
component of the hypermultiplet that is singlet under the diagonal subgroup
of the two $SU(2)$'s, thereby breaking $SU(2)_u\times SU(2)_v$ to this
diagonal $SU(2)$ subgroup. The non-perturbative description of this class of
vacua will be provided by choosing $f$ and $g$ such that a pair of 
coincident seven
branes parallel to the $v$ plane smoothly join a pair of coincident seven
branes parallel to the $u$ plane, thus giving just one pair of coincident
seven branes. Such configurations have been discussed in the context of
$F$-theory in \cite{VAFSIX}.

\bigskip

\noindent{\bf Acknowledgement}:  I wish to thank R. Gopakumar, D. Jatkar,
S. Kalyanarama and S. Mukhi for useful discussions and
J. Sonnenschein, S. Theisen and 
S. Yankielowicz for useful communications.


\begin{thebibliography}{99}

\bibitem{ORIENT}
A. Sagnotti, `Open Strings and their Symmetry Groups', Talk at
Cargese Summer Inst., 1987; \\
G. Pradisi and A. Sagnotti, Phys. Lett. {\bf B216} (1989) 59; \\
M. Bianchi, G. Pradisi and A. Sagnotti, Nucl. Phys. {\bf B376}
(1992) 365; \\
P. Horava, Nucl. Phys. {\bf B327} (1989) 461,
Phys. Lett. {\bf B231} (1989) 251.

\bibitem{DBRANE}
J. Dai, R. Leigh, and J. Polchinski, Mod. Phys. Lett. 
{\bf A4} (1989) 2073; \\
R. Leigh, Mod. Phys. Lett. {\bf A4} (1989) 2767; \\
J. Polchinski, Phys. Rev. {\bf D50} (1994) 6041 [hep-th/9407031].
 
\bibitem{WITTEND}
E. Witten, Nucl. Phys. {\bf B443} (1995) 85 [hep-th/9503124].

\bibitem{POLCHI}
J. Polchinski, Phys. Rev. Lett. {\bf 75} (1995) 4724 [hep-th/9510017].

\bibitem{SENF}
A. Sen, hep-th/9605150.

\bibitem{GIMPOL}
E. Gimon and J. Polchinski, Phys. Rev. {\bf D54} (1996) 1667 
[hep-th/9601038]. 

\bibitem{BIAN}
M. Bianchi and A. Sagnotti, Nucl. Phys. {\bf B361} (1991) 519; \\
C. Angelantonj, M. Bianchi, G. Pradisi, A. Sagnotti and Y. Stanev,
Phys. Lett. {\bf B385} (1996) 96 [hep-th/9606169].
Phys. Lett. {\bf B387} (1996) 743 [hep-th/9607229].

\bibitem{COSH}
O. Aharony, J. Sonnenschein, S. Yankielowicz and 
S. Theisen, hep-th/9611222.

\bibitem{BDS}
T. Banks, M. Douglas and N. Seiberg, hep-th/9605199.

\bibitem{POLNEW}
J. Polchinski, hep-th/9606165.

\bibitem{BLUZAF}
J. Blum and A. Zaffaroni, hep-th/9607019.

\bibitem{DABPAR}
A. Dabholkar and J. Park, hep-th/9607041.

\bibitem{GIMJOH}
E. Gimon and C. Johnson, hep-th/9606176.

\bibitem{GOPMUK}
R. Gopakumar and S. Mukhi, hep-th/9607057.

\bibitem{WITSIX}
M. Berkooz, R. Leigh, J. Polchinski, J. Schwarz, N. Seiberg and E. Witten,
hep-th/9605184.

\bibitem{SEIWIT}
N. Seiberg and E. Witten, Nucl. Phys. {\bf B431} (1994) 484 
[hep-th/9408099].

\bibitem{FTHEORY}
C. Vafa, Nucl. Phys. {\bf B469} (1996) 403 [hep-th/9602022]; \\
D. Morrison and C. Vafa, Nucl. Phys. {\bf B473} (1996) 74;
[hep-th/9602114]; hep-th/9603161.

\bibitem{VAFSIX}
M. Bershadsky, K. Intrilligator, S. Kachru, D. Morrison, V. Sadov
and C. Vafa, hep-th/9605200.

\bibitem{OTHERS}
M. Bershadsky and A. Johansen, hep-th/9610111; \\
O. Aharony, S. Kachru and E. Silverstein, hep-th/9610205.

\bibitem{SEILAT}
N. Seiberg, hep-th/9606017, hep-th/9608111; \\
N. Seiberg and E. Witten, hep-th/9607163; \\
D. Morrison and N. Seiberg, hep-th/9609070.

\bibitem{DOULAT}
M. Douglas, S. Katz and C. Vafa, hep-th/9609071.

\bibitem{ASPINK3}
P. Aspinwall, Phys. Lett. {\bf B357} (1995) 329 [hep-th/9507012].

\bibitem{SENBPS}
A. Sen, hep-th/9608005.

\end{thebibliography}
\end{document}